\documentclass[twocolumn,portuguese]{IEEEtran}
\usepackage[utf8]{inputenc}
\usepackage{graphicx}
\usepackage{babel}
\usepackage{subcaption}
\usepackage{color}
\usepackage{soulutf8}
\usepackage{hyperref}
\usepackage[square,numbers]{natbib}

\providecommand{\keywords}[1]
{
  \small	
  \textbf{\textit{Palavras chave---}} #1
}
  
\title{Enlaces de rádio de longa distância utilizando a banda de HF}

\author{Rafael~Diniz,~Mylène~C.~Q.~Farias\\~\IEEEmembership{Universidade~de~Brasília\\}
}
\begin{document}

\maketitle

\begin{abstract}
O interesse no uso da banda de Ondas Curtas (ou HF, do inglês High Frequency) em telecomunicação aumentou significativamente na última década, principalmente devido ao desenvolvimento de novos padrões para telecomunicação militar em HF, assim como com a expansão da radiodifusão digital na faixa HF. Mais especificamente, este novos padrões permitem implementar enlaces de centenas ou milhares de quilômetros a um baixo custo, o que permite a sua utilização mais ampla. No Brasil, este tipo de comunicação pode ser utilizado em regiões remotas ou de difícil acesso, como a região amazônica. Além das evoluções de tecnologias relativas à camada física, tem havido um grande desenvolvimento de técnicas que utilizam algoritmos de aprendizado de máquina para codificação de sinais (sons e imagens). Acredita-se que todos estes avanços irão possibilitar a utilização da banda de HF para transmissões em locais sem infraestrutura de telecomunicações. Este trabalho apresenta aplicações recentes de enlaces digitais de rádio em HF no Brasil, descrevendo os desafios  presentes para o desenvolvimento de sistemas de telecomunicação na faixa HF. \\
Artigo publicado na SET eXperience 2020. Enlace da apresentação em vídeo: \url{https://youtu.be/shkyEcHl7Fk}.
\end{abstract}


\section{Introdução}
O interesse no uso da banda de Ondas Curtas (ou HF, do inglês \textit{High Frequency}) em telecomunicação aumentou significativamente na última década, principalmente devido ao desenvolvimento de novos padrões para telecomunicação militar em HF, assim como com a expansão da radiodifusão digital na faixa HF (Ondas Curtas). Mais especificamente, este novos padrões permitem implementar enlaces de centenas ou milhares de quilômetros a um baixo custo, o que permite a sua utilização mais ampla.  Recentemente, Witvliet e Alsina-Pagès~\cite{witvliet2017radio} publicaram um estudo do estado da arte sobre comunicação na faixa HF. Neste documento, eles discutem o ``renascimento'' da banda HF com foco no modo de transmissão NVIS (do inglês \textit{Near Vertical Incidence Skywave}), que se caracteriza pelo padrão de irradiação vertical da antena na direção do céu. Witvliet e Alsina-Pagès apresentam os padrões militares de telecomunicação em HF e a radiodifusão digital em HF. 

\begin{figure*}[!ht]
    \centering
     \includegraphics[scale=0.3]{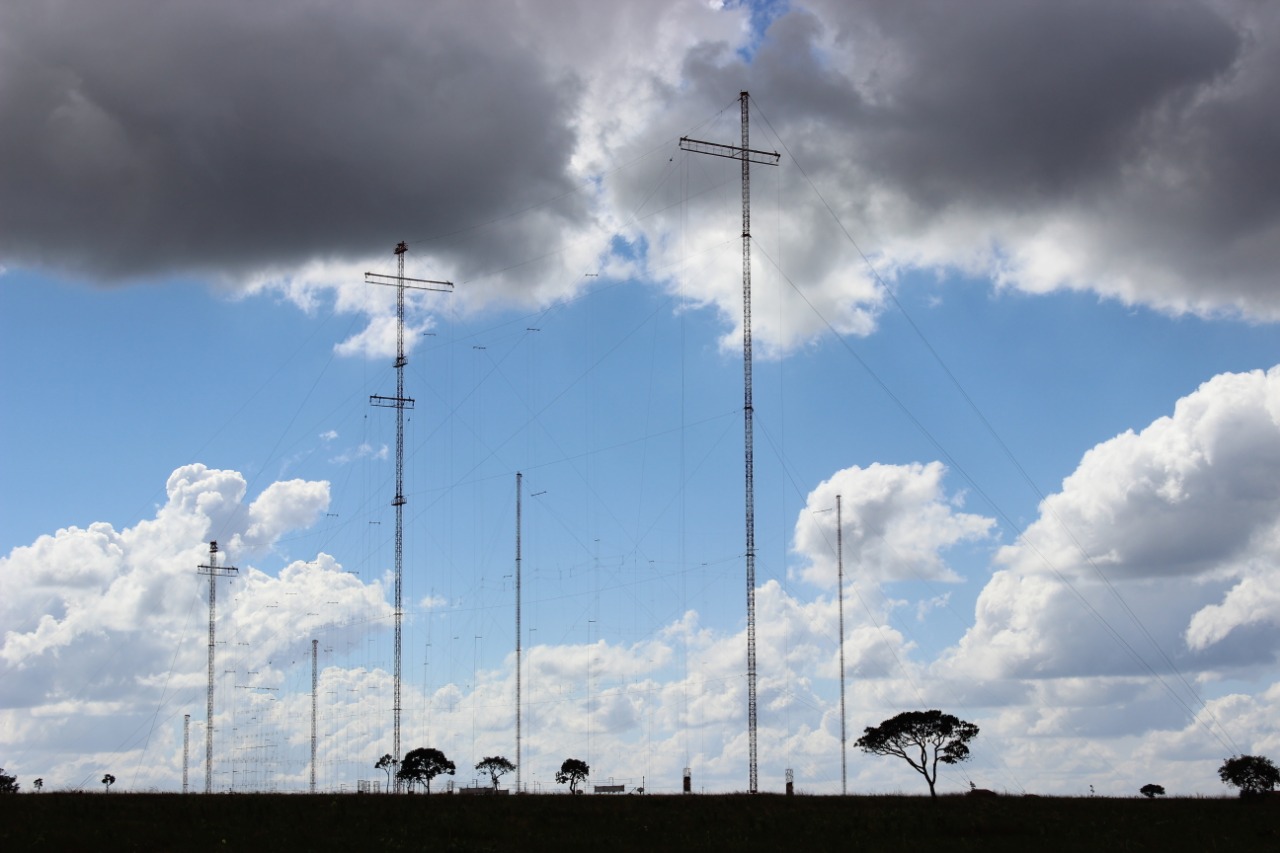}
    \caption{Uma das antenas para transmissão em HF no Parque do Rodeador, instalação da EBC localizada no Distrito Federal.}
    \label{fig:ebc}
\end{figure*}

Muitas localidades no Brasil, como por exemplo várias partes da região amazônica, estão localizadas a uma grande distância de centros urbanos. Este isolamento dificulta o acesso aos meios de telecomunicações básicos, como telefonia ou internet.  Em particular, nas regiões da floresta amazônica, a densa vegetação e os diversos rios da floresta dificultam a instalação de fibras ópticas ou altas torres para enlaces de rádio com linha de visada.  Neste cenário, a utilização de sistemas de telecomunicação digitais em HF permitiriam prover serviços de comunicações às populações locais. Infelizmente, a telecomunicação digital na faixa de HF tem sido utilizada quase que exclusivamente para aplicações militares. Com o objetivo de avaliar as possibilidades de prestação de serviços de comunicação na faixa HF para aplicações civis na região amazônica, a Empresa Brasil de Comunicação (EBC), em uma parceria entre a Universidade de Brasília e o Ministério das Comunicações, está testando o sistema de radiodifusão digital Digital Radio Mondiale (DRM)~\cite{diniz2011sistema} em HF na frequência de 11.910~kHz. A Figura~\ref{fig:ebc} apresenta uma das antenas de transmissão em HF da EBC.

\begin{figure*}[!ht]
    \centering
     \includegraphics[scale=0.4]{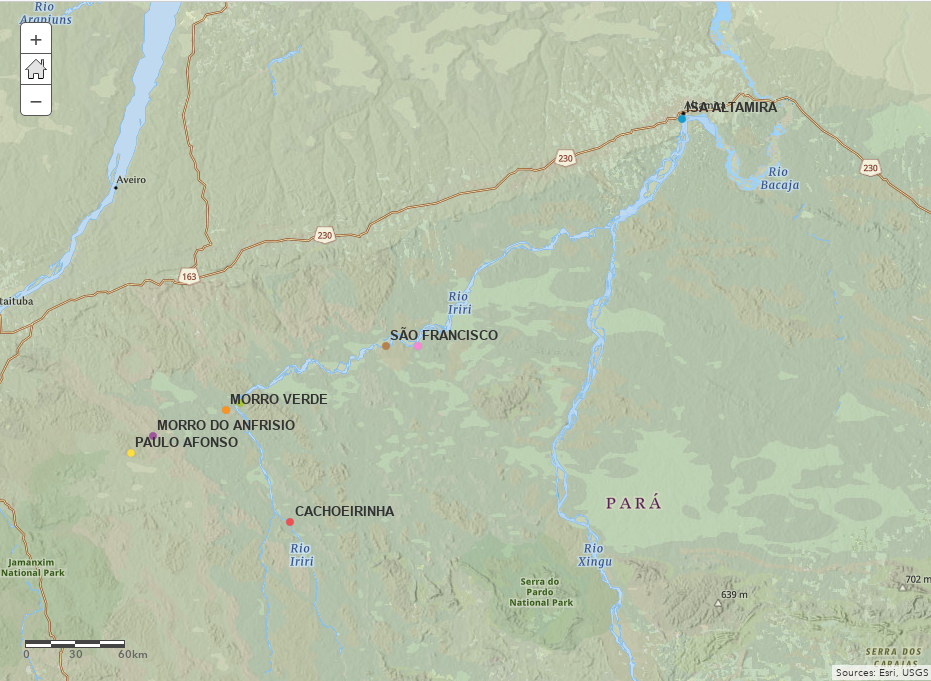}
    \caption{Mapa ilustrando as localizações das localidades onde foram instaladas estações HF no Pará.}
    \label{fig:mapa}
\end{figure*}

Neste artigo, apresentamos um estudo sobre três projetos para utilização de enlaces de longa distância na faixa HF, com objetivo de atender populações das comunidades da floresta, das pequenas cidades da região amazônica ou em outras partes do mundo. Dois dos projetos descritos neste trabalho estão sendo implementados na região amazônica: um no Acre, no município de Marechal Thaumaturgo, e outro no Pará, em Altamira. O terceiro projeto foi implementado no México, mais especificamente no estado de Oaxaca. Em todos os três projetos descritos foram utilizados transceptores Single Side Band (SSB)  ``de prateleira'' para a faixa HF e enlaces estabelecidos através de propagação celeste (em inglês, Skywave). Vale salientar que no Skywave o caminho do sinal entre o emissor e o receptor se dá através de pelo menos uma reflexão na ionosfera. No caso do projeto realizado no Acre, que teve início em 2015~\cite{caminati2015rede}, as estações de HF foram instaladas na Reserva Extrativista do Alto Juruá. Este projeto tem o objetivo de estabelecer uma comunicação da comunidade local com o centro urbano do município mais próximo, que se encontra a aproximadamente 70~km da reserva. No caso do projeto implementado no México, em 2018, mensagens SMS e gravações de voz foram transmitidas via HF. Neste caso, a comunidade é localizada no interior do estado de Oaxaca, sendo cercada por montanhas. Um enlace foi estabelecido entre esta comunidade e a capital do estado. Finalmente, no projeto implementado no Pará em 2019  algumas comunidades na região da Terra do Meio~\footnote{Terra do Meio: \\ \url{https://en.wikipedia.org/wiki/Terra_do_Meio_Ecological_Station}} foram conectadas a Altamira. As distâncias neste caso são  entre 300 e 500km,  como ilustrado na imagem da Figura~\ref{fig:mapa}.

Todos os projetos descritos partem da premissa que a transmissão na faixa HF é mais adequada a este tipo de região remota devido às suas caraterísticas técnicas e ao seu baixo custo, que é bem inferior ao que seria necessário para instalação de fibras ópticas, alta torres para enlaces de rádio ou mesmo enlaces de satélites.  O protótipo utilizado nestes projetos de transmissão na faixa HF consiste de uma antena de dipolo de $1/4 \lambda$ feita com fio de cobre, instalada em V invertido e conectada a um rádio transceptor para HF. A modulação digital e outros elementos da pilha de rede rodam num computador com Linux que é conectado ao rádio HF. Esta solução simples permite enlaces servindo a distâncias de várias centenas de quilômetros. É importante salientar que a banda de HF foi escolhida por permitir a propagação celeste~\cite{davies1965ionospheric}, na qual o sinal deixando o planeta reflete nas camadas E e F da ionosfera de volta à superfície.  Entretanto, o uso de propagação celeste traz desafios distintos de outros modos de propagação, devido aos efeitos de desvanecimento, multi-percurso e espalhamento Doppler. Desta forma, uma pilha distinta de protocolos da Internet ou de redes de telefonia, otimizada para um canal de comunicação intermitente e com baixa taxa de transmissão de bits, é mais adequada para a transmissão na faixa  HF.

\section{Telecomunicações em HF atualmente}

Como já mencionado, no Brasil atualmente existem poucos projetos civis que utilizem técnicas de telecomunicação digital na faixa de HF. Um dos poucos exemplos é o projeto Fonias Juruá~\cite{caminati2015rede} que teve início em 2015. Neste projeto, foi desenvolvido um sistema para transmissão de arquivos baseada no sistema de radiodifusão DRM. Para isto,  foi implementada  uma adaptação do DRM para estreitar banda do sinal, de forma a permitir a utilização de um rádio HF SSB comum, com banda passante em torno de 3 kHz. Os arquivos são transmitidos no modo de carrossel de dados, onde um arquivo é transmitido em um laço contínuo de repetições. No Peru, na década de 2000, o projeto EHAS~\cite{seoane2009ehas} utilizou, dentre outras tecnologias, enlaces digitais em HF para conectar postos de saúde na Amazônia peruana. Vale lembrar que o uso da banda de HF para rádio fonia SSB analógica na Amazônia é altamente difundido, sendo uma das regiões com mais elevado uso de rádio fonia em HF para comunicação civil no mundo.

Em 2018, foi realizada uma competição promovida pela Mozilla Foundation~\footnote{Mozilla Wireless Challenge: \url{https://wirelesschallenge.mozilla.org/}} para tratar sobre desafios de redes de comunicação em situações de desastres e emergências. A organização não governalmental (ONG) Rhizomatica participou da competição com o projeto HERMES e ganhou o primeiro prêmio. Para a ocasião foi realizada uma demonstração de um enlace em HF conectando uma comunidade em um vale nas montanhas de Oaxaca à capital do estado mexicano de mesmo nome. Nesta demonstração, mensagens SMS (short message service) e gravações de voz foram transmitidas na faixa HF, sendo que os usuários conectavam-se a uma rede local através de uma rede GSM provida por uma pequena estação rádio base GSM~\footnote{Vídeo da demonstração no México: \url{https://vimeo.com/398331581} .}. Em 2019, o sistema foi melhorado, permitindo a sua primeira instalação em campo, que foi feita no Pará. Nesta melhoria, o sistema foi modificado para que os usuários pudessem acessar os serviços providos pela rede em HF através de uma rede WiFi local, o que permitiu, por exemplo, enviar e receber uma foto através do celular.

\begin{figure*}[!ht]
    \centering
    \includegraphics[scale=0.5]{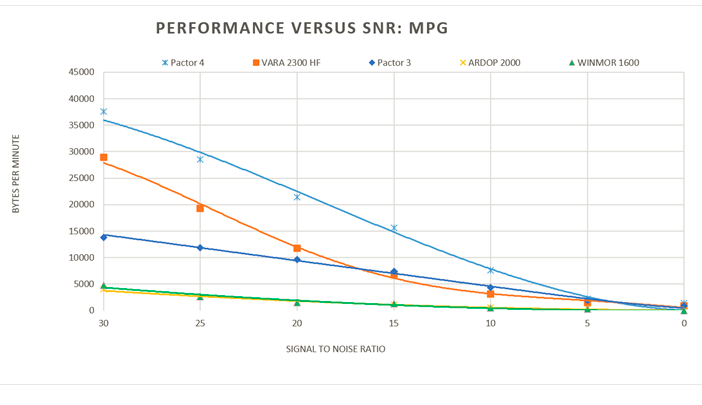}
    \caption{Comparativo da taxa de transmissão de modems na faixa HF.}
    \label{fig:modem}
\end{figure*}

O projeto HERMES~\footnote{High-frequency Emergency and Rural Multimedia Exchange System: \url{https://www.rhizomatica.org/hermes/}}, utiliza um modem baseado em software de nome \textit{Amateur Radio Digital Open Protocol} (ARDOP)~
\footnote{\url{http://www.winlink.org/content/ardop_overview}}. Diferentemente do DRM, que é feito para radiodifusão, o ARDOP foi desenvolvido por rádio-amadores para troca de dados de forma bi-direcional. Como já mencionado, o uso de uma pilha de rede baseada no \textit{Internet Protocol} (IP) apresenta grandes desafios para uso na faixa HF, como descrito por Elsadig~\cite{elsadig2017design}. Desta forma, protocolos que suportam o paradigma \textit{store-and-forward}, permitindo a transferência de arquivos e mensagens de forma assíncrona e agrupada, são de grande importância para telecomunicação na faixa HF. Em particular, o protocolo utilizado pelo  projeto HERMES~\footnote{Código fonte do projeto HERMES: \url{http://github.com/DigitalHERMES}} é o \textit{Unix-to-Unix Copy} (UUCP)~\cite{nowitz1978uucp}. O UUCP,  foi desenvolvido pela Bell Labs no final da década de 70, tendo como objetivo a transmissão de arquivos e a execução remota de comandos entre computadores Unix conectados via linha telefônica. O UUCP permite a implementação de serviços como e-mail, \textit{news}, troca de arquivos e execução de comandos remotos de forma assíncrona. Grandes atrasos ou mesmo interrupções temporárias de enlace não impedem o funcionamento correto de serviços baseados no UUCP.  

Rick Muething, criador do ARDOP, realizou uma análise das opções de modem~\footnote{Modem neste contexto significa solução de transmissão digital que provê um canal de comunicação sem perda para a aplicação.} disponíveis para uso civil, considerando uma banda passante limitada a 3 kHz (devido à limitação dos rádios HF). Na Figura~\ref{fig:modem} é apresentado um gráfico de resultados deste trabalho, no qual  distintos modems HF são comparados. O eixo horizontal do gráfico corresponde à relação sinal ruído (SNR), enquanto que o eixo vertical corresponde à taxa de transmissão em bytes por minuto para um canal MPG (MultiPath Good, assim como definido em recomendação da ITU~\cite{barclay1997hf}). Os modems comparados são: o Pactor 4 e o Pactor 3, ambos proprietários e implementados em hardware pelo fabricante,  O VARA, que é um software para Windows desenvolvido por um rádio amador, o WINMOR, que é um software grátis para Windows, e o ARDOP, que tem o código fonte aberto e é compatível com Linux e Windows.

Diversas limitações relativas ao baixo desempenho do modem ARDOP levaram a distintos esforços para o desenvolvimento de um novo modem aberto baseado em software. Opções para permitir transmissões tanto em portadora única, com \textit{Frequency-shift Keying} (FSK), como em múltiplas portadoras, com \textit{Orthogonal Frequency Division Multiplexing} (OFDM) e a utilização de códigos corretores de erros \textit{Low-density parity-check code} (LDPC) estão sendo consideradas pelo Rhizomatica e rádio amadores como David Rowe\footnote{Desenvolvimentos de modem para dados do David Rowe: \url{https://bit.ly/3lW7Vgq}}. O uso de sistemas de portadora única, por exemplo, é justificado devido à opção por esse tipo de modulação pelos padrões militares para transceptores HF da Organização do Tratado do Atlântico Norte (OTAN), conforme detalhado pelo Professor Eric E. Johnson em publicação recente~\cite{johnson2020}.

\begin{figure*}[!ht]
    \centering
    \begin{subfigure}[b]{0.45\linewidth}
        \centering
        \includegraphics[width=\linewidth]{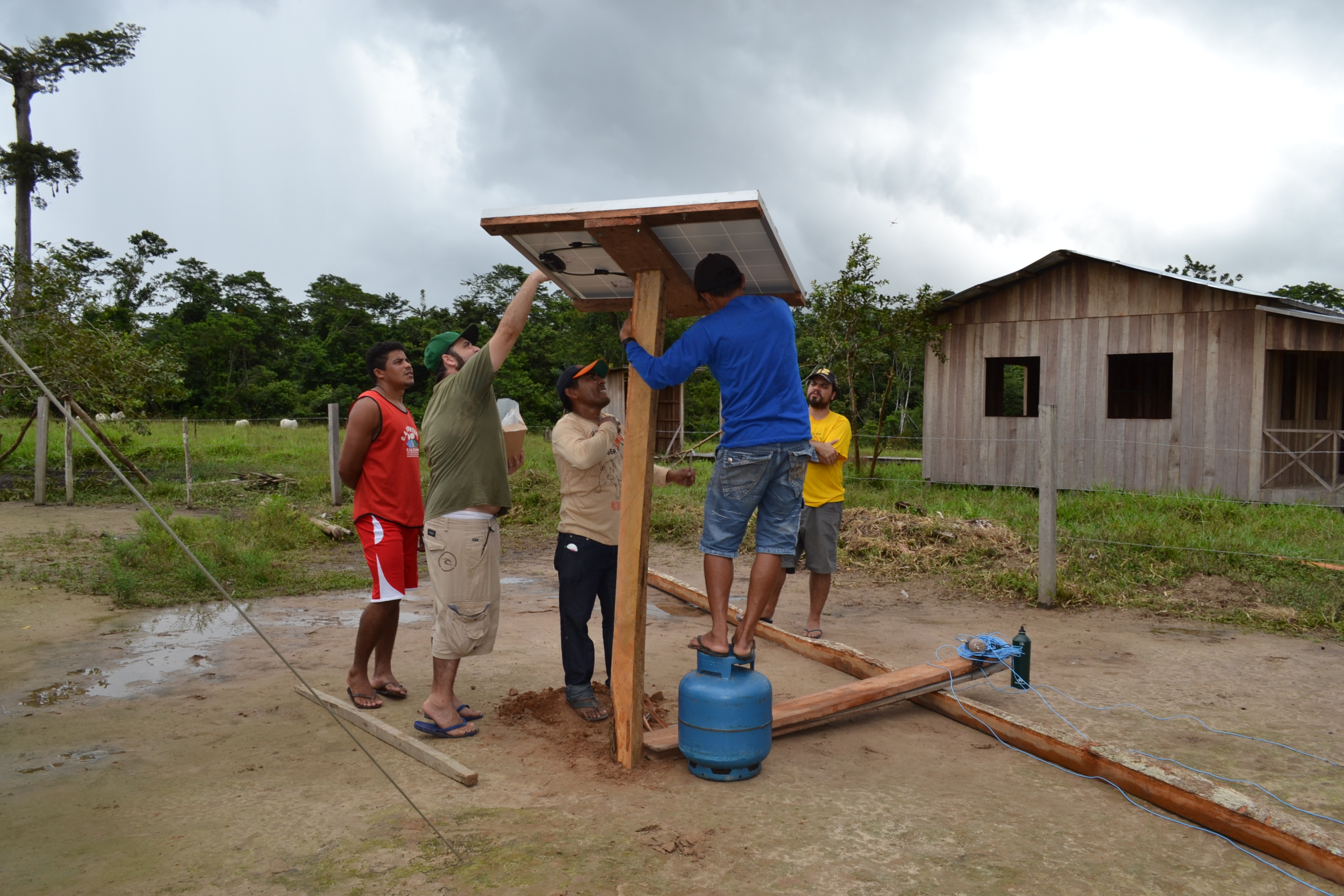}
        \subcaption{Painel Solar.}
        \label{fig:cenariob}
    \end{subfigure}
    \begin{subfigure}[b]{0.45\linewidth}
        \centering
        \includegraphics[width=\linewidth]{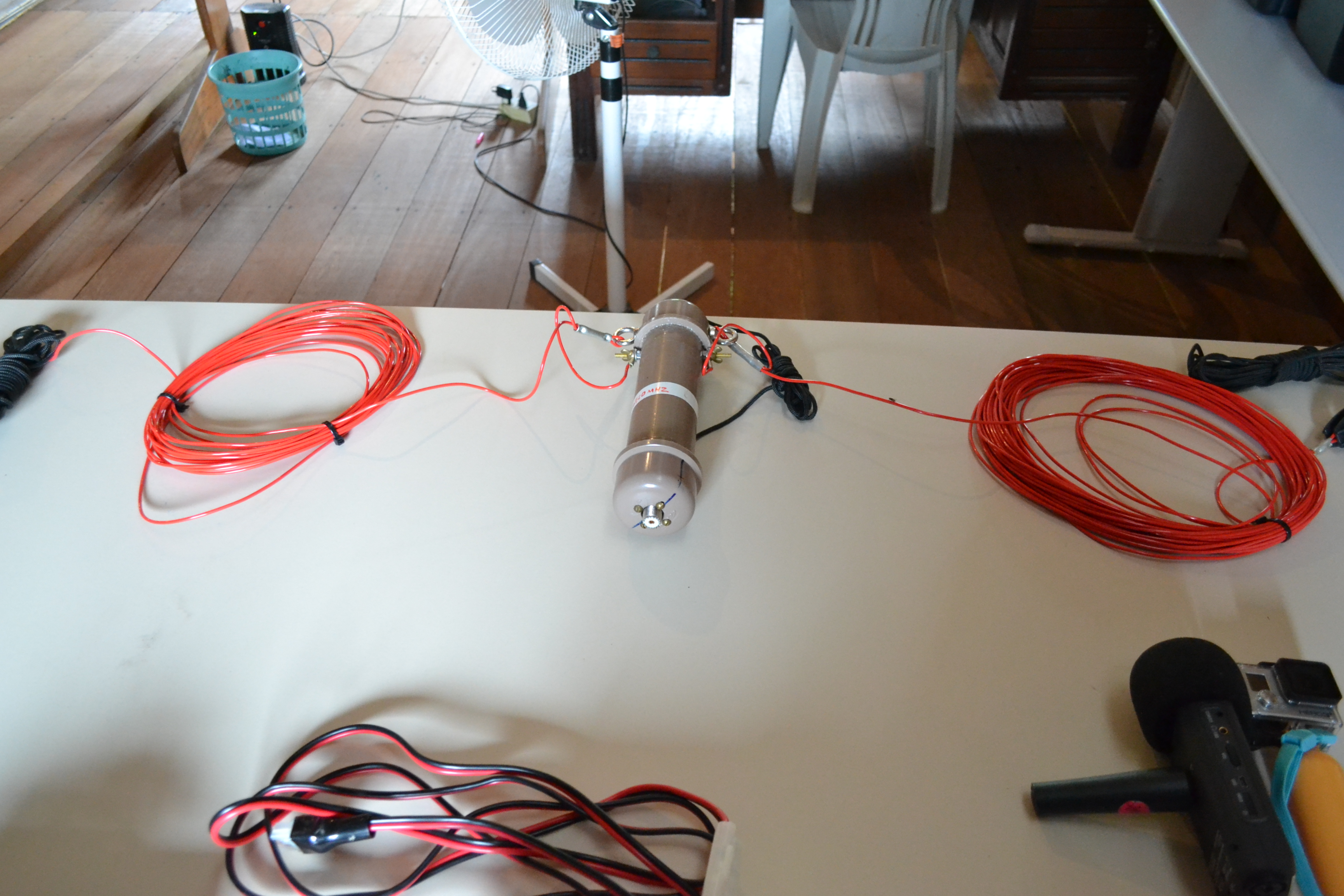}
        \subcaption{Antena de transmissão.}
        \label{fig:cenarioa}
    \end{subfigure}
    \begin{subfigure}[b]{0.45\linewidth}
        \centering
        \includegraphics[width=\linewidth]{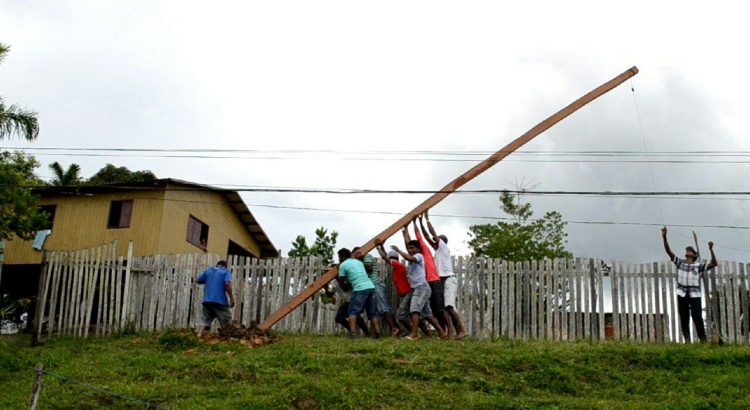}
        \subcaption{Montagem da antena.}
        \label{fig:cenarioc}
    \end{subfigure}
    \begin{subfigure}[b]{0.45\linewidth}
        \centering
        \includegraphics[width=\linewidth]{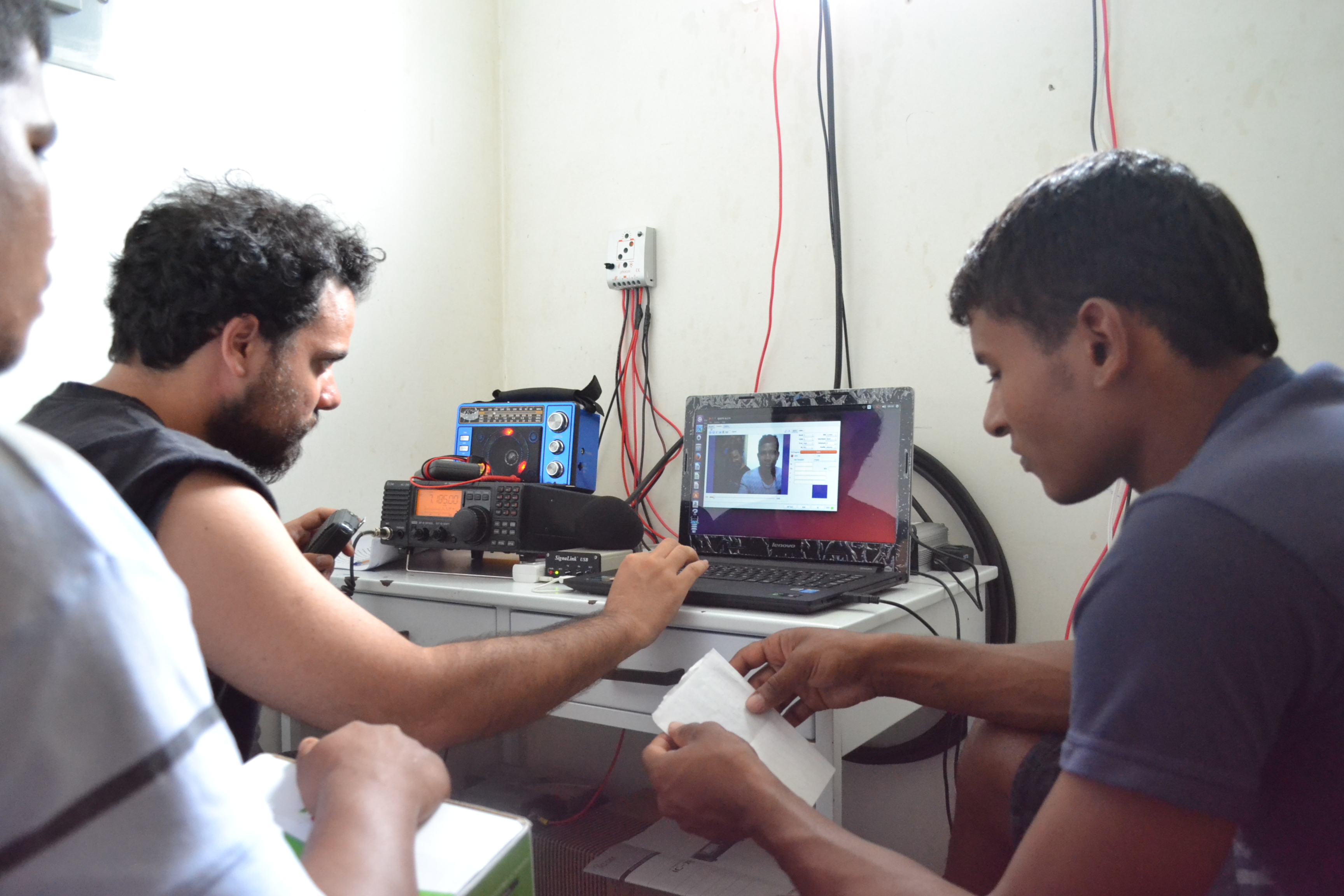}
        \subcaption{Configuração de equipamentos.}
        \label{fig:cenariod}
    \end{subfigure}
    \caption{Fotos de uma localidade no Acre no contexto do projeto Fonias Juruá.}
\end{figure*}

Um nó de rede na banda de HF na região amazônica, por exemplo, tipicamente envolve a instalação de um painel solar (Fig. ~\ref{fig:cenarioa}), um regulador de carga, bateria, postes para sustentação da antena (Fig. ~\ref{fig:cenarioc}), a antena (Fig.~\ref{fig:cenariob}), um cabo para conectar o transceptor à antena, e o transceptor para a faixa HF conectado a um computador (Fig.~\ref{fig:cenariod}).  Para contornar a baixíssima taxa de transmissão em um canal HF com a tecnologia disponível, uma busca por codificadores de fonte mais eficientes também vem sendo conduzida. Por exemplo, no experimento realizado para envio  de mensagens de áudio via HF no México foi utilizado o codificador de voz Codec2~\cite{rowe1997techniques}. Foi considerada uma taxa de 1.300 bit/s, que foi escolhida após testes subjetivos de qualidade (o modo de 700 bit/s não apresentou qualidade satisfatória e foi rejeitado). Em 2019, uma evolução do Codec2, com o uso de aprendizado de máquina~\cite{erhardt2019open}, permitiu o desenvolvimento de uma versão de 450 bit/s. 

Outra proposta recente consiste na aplicação de uma rede neural computacional, a LPCNet~\cite{valin2019lpcnet}, para codificação de voz~\cite{valin2019real} a uma taxa de 1.600 bit/s. Este sistema propicia uma qualidade muito superior quando comparado a codificadores que não utilizam técnicas de aprendizado de máquina. Em outras palavras, o sistema produz uma melhor qualidade subjetiva para  uma mesma taxa de bits, mas por outro lado tem uma maior complexidade computacional. A mesma tendência se observa para codificação de imagens em taxas muito baixas, com várias pesquisas sendo  atualmente desenvolvidas nesta linha. Por exemplo, o codificador HiFiC~\cite{mentzer2020high}  apresenta grandes ganhos de compressão quando comparado aos padrões mais recentes para codificação de imagem. 

\section{Considerações Finais}

Existe uma grande assimetria tecnológica entre sistemas de telecomunicações HF para fins militares~\cite{johnson2016wideband} e os para fins civis. Os sistemas militares utilizam larguras de banda maiores que o tradicional canal de 3~kHz da banda de HF, atingindo taxas de transmissão superiores a 100~kbit/s e larguras de banda de até 48~kHz. Já os sistemas civis para faixa HF, além da função de telefonia analógica SSB, tipicamente não dispõem de qualquer sistema para comunicação digital, estando limitados a larguras de banda passante ao redor de 3kHz, o que é suficiente para voz. No caso de transmissão de dados digitais, em um canal de HF típico e um enlace de 600~km de distância é possível obter uma taxa de transmissão em torno de 1~kbit/s em um canal de 3~kHz, o que não é muito. Desta forma, concluímos que o desenvolvimento de transceptores HF  de banda larga e de uma pilha de rede otimizada para uso civil em HF pode impulsionar a adoção da banda HF para telecomunicação digital em aplicações civis, ao permitir taxas de transmissão mais altas.

O presente trabalho foi realizado com apoio da Coordenação de Aperfeiçoamento de Pessoal de Nível Superior - Brasil (CAPES).






\bibliographystyle{ACM-Reference-Format}
\bibliography{_main}


\begin{thebibliography}{15}


\ifx \showCODEN    \undefined \def \showCODEN     #1{\unskip}     \fi
\ifx \showDOI      \undefined \def \showDOI       #1{#1}\fi
\ifx \showISBNx    \undefined \def \showISBNx     #1{\unskip}     \fi
\ifx \showISBNxiii \undefined \def \showISBNxiii  #1{\unskip}     \fi
\ifx \showISSN     \undefined \def \showISSN      #1{\unskip}     \fi
\ifx \showLCCN     \undefined \def \showLCCN      #1{\unskip}     \fi
\ifx \shownote     \undefined \def \shownote      #1{#1}          \fi
\ifx \showarticletitle \undefined \def \showarticletitle #1{#1}   \fi
\ifx \showURL      \undefined \def \showURL       {\relax}        \fi
\providecommand\bibfield[2]{#2}
\providecommand\bibinfo[2]{#2}
\providecommand\natexlab[1]{#1}
\providecommand\showeprint[2][]{arXiv:#2}

\bibitem[\protect\citeauthoryear{Barclay}{Barclay}{1997}]%
        {barclay1997hf}
\bibfield{author}{\bibinfo{person}{LW Barclay}.}
  \bibinfo{year}{1997}\natexlab{}.
\newblock \showarticletitle{HF simulator standards within the ITU}. In
  \bibinfo{booktitle}{\emph{Seventh International Conference on HF Radio
  Systems and Techniques}}. IET, \bibinfo{pages}{359--361}.
\newblock


\bibitem[\protect\citeauthoryear{Caminati and Diniz}{Caminati and
  Diniz}{2015}]%
        {caminati2015rede}
\bibfield{author}{\bibinfo{person}{Francisco~Antunes Caminati} {and}
  \bibinfo{person}{Rafael Diniz}.} \bibinfo{year}{2015}\natexlab{}.
\newblock \showarticletitle{Rede Fonias Juru{\'a}: tecnologia, territ{\'o}rio e
  cultura para al{\'e}m da {\'u}ltima milha da Rede Mundial}.
\newblock \bibinfo{journal}{\emph{Anais do III Encontro Brasileiro de Pesquisa
  em Cultura [Recurso Eletr{\^o}nico]. Crato/CE: Universidade Federal do
  Cariri}}  \bibinfo{volume}{1} (\bibinfo{year}{2015}),
  \bibinfo{pages}{84--93}.
\newblock


\bibitem[\protect\citeauthoryear{Davies}{Davies}{1965}]%
        {davies1965ionospheric}
\bibfield{author}{\bibinfo{person}{Kenneth Davies}.}
  \bibinfo{year}{1965}\natexlab{}.
\newblock \bibinfo{booktitle}{\emph{Ionospheric radio propagation}}.
  Vol.~\bibinfo{volume}{80}.
\newblock \bibinfo{publisher}{US Department of Commerce, National Bureau of
  Standards}.
\newblock


\bibitem[\protect\citeauthoryear{Diniz}{Diniz}{2011}]%
        {diniz2011sistema}
\bibfield{author}{\bibinfo{person}{Rafael Diniz}.}
  \bibinfo{year}{2011}\natexlab{}.
\newblock \showarticletitle{O sistema Digital Radio Mondiale no contexto de
  escolha da norma técnica para o Sistema Brasileiro de Rádio Digital}.
\newblock \bibinfo{journal}{\emph{Revista de Radiodifusão}}
  \bibinfo{volume}{5}, \bibinfo{number}{5} (\bibinfo{year}{2011}).
\newblock


\bibitem[\protect\citeauthoryear{Elsadig}{Elsadig}{2017}]%
        {elsadig2017design}
\bibfield{author}{\bibinfo{person}{Elsadig~Gamaleldeen Elsadig}.}
  \bibinfo{year}{2017}\natexlab{}.
\newblock \showarticletitle{Design of IP over HF channels under sudan
  ionospheric transmissions conditions}. In \bibinfo{booktitle}{\emph{2017
  Joint International Conference on Information and Communication Technologies
  for Education and Training and International Conference on Computing in
  Arabic (ICCA-TICET)}}. IEEE, \bibinfo{pages}{1--7}.
\newblock


\bibitem[\protect\citeauthoryear{Erhardt, Kurin, Lurz, Weigel, and
  Koelpin}{Erhardt et~al\mbox{.}}{2019}]%
        {erhardt2019open}
\bibfield{author}{\bibinfo{person}{Stefan Erhardt}, \bibinfo{person}{Thomas
  Kurin}, \bibinfo{person}{Fabian Lurz}, \bibinfo{person}{Robert Weigel}, {and}
  \bibinfo{person}{Alexander Koelpin}.} \bibinfo{year}{2019}\natexlab{}.
\newblock \showarticletitle{An Open-Source Speech Codec at 450 bit/s with
  Pseudo-Wideband Mode}. In \bibinfo{booktitle}{\emph{2019 49th European
  Microwave Conference (EuMC)}}. IEEE, \bibinfo{pages}{1048--1051}.
\newblock


\bibitem[\protect\citeauthoryear{Johnson}{Johnson}{2016}]%
        {johnson2016wideband}
\bibfield{author}{\bibinfo{person}{Eric~E Johnson}.}
  \bibinfo{year}{2016}\natexlab{}.
\newblock \showarticletitle{Wideband ALE--the next generation of HF}. In
  \bibinfo{booktitle}{\emph{2016 Nordic HF Conference, HF}},
  Vol.~\bibinfo{volume}{16}.
\newblock


\bibitem[\protect\citeauthoryear{Johnson}{Johnson}{2020}]%
        {johnson2020}
\bibfield{author}{\bibinfo{person}{Eric~E Johnson}.}
  \bibinfo{year}{2020}\natexlab{}.
\newblock \showarticletitle{Wideband Steps Up to Fill the Gap, 4G high
  frequency radios supplement satellite communications}. In
  \bibinfo{booktitle}{\emph{AFCEA Signal Magazine}}, Vol.~\bibinfo{volume}{4}.
  \bibinfo{publisher}{AFCEA Press}, \bibinfo{pages}{36--39}.
\newblock


\bibitem[\protect\citeauthoryear{Mentzer, Toderici, Tschannen, and
  Agustsson}{Mentzer et~al\mbox{.}}{2020}]%
        {mentzer2020high}
\bibfield{author}{\bibinfo{person}{Fabian Mentzer}, \bibinfo{person}{George
  Toderici}, \bibinfo{person}{Michael Tschannen}, {and}
  \bibinfo{person}{Eirikur Agustsson}.} \bibinfo{year}{2020}\natexlab{}.
\newblock \showarticletitle{High-Fidelity Generative Image Compression}.
\newblock \bibinfo{journal}{\emph{arXiv preprint arXiv:2006.09965}}
  (\bibinfo{year}{2020}).
\newblock


\bibitem[\protect\citeauthoryear{Nowitz}{Nowitz}{1978}]%
        {nowitz1978uucp}
\bibfield{author}{\bibinfo{person}{DA Nowitz}.}
  \bibinfo{year}{1978}\natexlab{}.
\newblock \bibinfo{booktitle}{\emph{Uucp implementation description}}.
  Vol.~\bibinfo{volume}{2}.
\newblock \bibinfo{publisher}{Bell Laboratories. In UNIX Programmer’s
  Manual}.
\newblock


\bibitem[\protect\citeauthoryear{Rowe}{Rowe}{1997}]%
        {rowe1997techniques}
\bibfield{author}{\bibinfo{person}{David~Grant Rowe}.}
  \bibinfo{year}{1997}\natexlab{}.
\newblock \emph{\bibinfo{title}{Techniques for harmonic sinusoidal coding}}.
\newblock \bibinfo{thesistype}{Ph.D. Dissertation}. \bibinfo{school}{University
  of South Australia}.
\newblock


\bibitem[\protect\citeauthoryear{Seoane~Pascual, S{\'a}nchez~Sala,
  Villarroel~Ortega, Mart{\'\i}nez~Fern{\'a}ndez, S{\'a}ez~Torres,
  et~al\mbox{.}}{Seoane~Pascual et~al\mbox{.}}{2009}]%
        {seoane2009ehas}
\bibfield{author}{\bibinfo{person}{Joaqu{\'\i}n Seoane~Pascual},
  \bibinfo{person}{Arnau S{\'a}nchez~Sala}, \bibinfo{person}{Valent{\'\i}n
  Villarroel~Ortega}, \bibinfo{person}{A Mart{\'\i}nez~Fern{\'a}ndez},
  \bibinfo{person}{Alberto S{\'a}ez~Torres}, {et~al\mbox{.}}}
  \bibinfo{year}{2009}\natexlab{}.
\newblock \showarticletitle{EHAS: programas libres para apoyar el sistema de
  salud en zonas aisladas de Am{\'e}rica Latina}.
\newblock  (\bibinfo{year}{2009}).
\newblock


\bibitem[\protect\citeauthoryear{Valin and Skoglund}{Valin and
  Skoglund}{2019a}]%
        {valin2019lpcnet}
\bibfield{author}{\bibinfo{person}{Jean-Marc Valin} {and} \bibinfo{person}{Jan
  Skoglund}.} \bibinfo{year}{2019}\natexlab{a}.
\newblock \showarticletitle{LPCNet: Improving neural speech synthesis through
  linear prediction}. In \bibinfo{booktitle}{\emph{ICASSP 2019-2019 IEEE
  International Conference on Acoustics, Speech and Signal Processing
  (ICASSP)}}. IEEE, \bibinfo{pages}{5891--5895}.
\newblock


\bibitem[\protect\citeauthoryear{Valin and Skoglund}{Valin and
  Skoglund}{2019b}]%
        {valin2019real}
\bibfield{author}{\bibinfo{person}{Jean-Marc Valin} {and} \bibinfo{person}{Jan
  Skoglund}.} \bibinfo{year}{2019}\natexlab{b}.
\newblock \showarticletitle{A real-time wideband neural vocoder at 1.6 kb/s
  using LPCNet}.
\newblock \bibinfo{journal}{\emph{arXiv preprint arXiv:1903.12087}}
  (\bibinfo{year}{2019}).
\newblock


\bibitem[\protect\citeauthoryear{Witvliet and Alsina-Pag{\`e}s}{Witvliet and
  Alsina-Pag{\`e}s}{2017}]%
        {witvliet2017radio}
\bibfield{author}{\bibinfo{person}{Ben~A Witvliet} {and}
  \bibinfo{person}{Rosa~Ma Alsina-Pag{\`e}s}.} \bibinfo{year}{2017}\natexlab{}.
\newblock \showarticletitle{Radio communication via Near Vertical Incidence
  Skywave propagation: an overview}.
\newblock \bibinfo{journal}{\emph{Telecommunication systems}}
  \bibinfo{volume}{66}, \bibinfo{number}{2} (\bibinfo{year}{2017}),
  \bibinfo{pages}{295--309}.
\newblock


\end{thebibliography}

\end{document}